\documentclass[twocolumn,
superscriptaddress,
 amsmath,amssymb,
 aps,
 pra, longbibliography
floatfix]{revtex4-2}
\usepackage{graphicx}
\usepackage{xcolor}
\usepackage[export]{adjustbox}
\usepackage{amsmath}
\usepackage{dcolumn}
\usepackage{physics}
\usepackage{bm}
\usepackage{hyperref}
\usepackage{soul}

\usepackage{transparent}

\newcommand{\mm}[1]{{\color{teal} #1}}

\newcommand{\floor}[1]{\lfloor #1 \rfloor}

\newcommand{\frah}[0]{\frac{1}{2}}
\newcommand{\hH}[0]{\hat{H}}

\newcommand{\hb}[1]{\hat{b}_{#1}}
\newcommand{\hbd}[1]{\hat{b}^\dagger_{#1}}
\newcommand{\hc}[1]{\hat{c}_{#1}}
\newcommand{\hcd}[1]{\hat{c}^\dagger_{#1}}
\newcommand{\hn}[1]{\hat{n}_{#1}}

\newcommand{\partj}[1]{\frac{\partial}{\partial #1}}

\definecolor{darkgreen}{rgb}{0.42, 0.66, 0.42}

\begin{document}

\title{
Super-Tonks-Girardeau Quench in the Extended Bose-Hubbard Model
}

\author{Maciej Marciniak}
\email{mmarciniak@cft.edu.pl}
\affiliation{Center for Theoretical Physics, Polish Academy of Sciences, Al. Lotnik\'{o}w 32/46, 02-668 Warsaw, Poland}

\author{Maciej Łebek}
\affiliation{Faculty of Physics, University of Warsaw, Pasteura 5, 02-093 Warsaw, Poland}

\author{Jakub Kopyciński}
\affiliation{Center for Theoretical Physics, Polish Academy of Sciences, Al. Lotnik\'{o}w 32/46, 02-668 Warsaw, Poland}

\author{Wojciech Górecki}
\affiliation{Faculty of Physics, University of Warsaw, Pasteura 5, 02-093 Warsaw, Poland}

\author{Rafał Ołdziejewski}
\affiliation{Max Planck Institute of Quantum Optics, 85748 Garching, Germany}

\author{Krzysztof Pawłowski}
\affiliation{Center for Theoretical Physics, Polish Academy of Sciences, Al. Lotnik\'{o}w 32/46, 02-668 Warsaw, Poland}

\date{\today}

\begin{abstract}

We investigate the effect of a quench from a one-dimensional gas with strong and repulsive local interactions to a strongly attractive one, known as the super-Tonks-Girardeau effect. By incorporating both an optical lattice and non-local interactions (specifically nearest-neighbor), we discover a previously unexplored phenomenon: the disruption of the state during the quench, but within a specific range of interactions. Our study employs the extended Bose-Hubbard model across various system sizes, starting with analytical results for two atoms and progressing to few-body systems using exact diagonalization, DMRG and TDVP methods. Finally, we use a numerical implementation of the local density approximation for a macroscopic number of atoms. Consistently, our findings unveil a region where the initially self-bound structure expands due to the super-Tonks-Girardeau quench. The fast evaporation provides a tool to characterize the phase diagram in state-of-art experiments exploring the physics of the extended Bose-Hubbard model.
\end{abstract}

\maketitle

\section{Introduction} \label{sec:introduction}
The studies of dynamics, particularly in one-dimensional configurations, generated by a quantum quench try to answer many fundamental questions related to the equilibration of the many-body system~\cite{Rigol2008,Dalessio}.
A typical quench scenario involves a rapid change of local interactions between particles. The system prepared in the ground state is driven far from equilibrium by a sudden change in the interaction strength.
After a long time, the system eventually relaxes and can be described using (generalized) Gibbs ensembles~\cite{Caux2012,Ilievski2015,Caux2013}.

A different behavior may be observed when bosons strongly repel each other. Therein, the system is in the ground state of the Tonks-Girardeau (TG) gas.~\cite{Girardeau1960, Kinoshita2004, Paredes2004}. Surprisingly, a quench to strong attraction does not cause violent dynamics or a collapse~\cite{Astrakharchik2005,Chen2010b}. Instead, the system reaches a stationary state called the super-Tonks-Girardeau (sTG) gas~\cite{Astrakharchik2005,Batchelor2005,Kormos2011}, whose properties resemble the ones of the pre-quench state.  Nonetheless, the sTG phase is more strongly correlated than the TG gas, and its equation of state maps to that of a hard-rod gas~\cite{Girardeau2010c} rather than ideal fermions. The difference between these phases, clearly visible in stiffness coefficient~\cite{Astrakharchik2005} and related to the breathing mode frequency, was demonstrated experimentally~\cite{Haller2009}. Similar physics was explored in one-dimensional gases for a slew of fermionic and bosonic models~\cite{Batchelor2005,Lieb1963b,Franchini2017,Guan2010,Guan2013,Chen2010a,Yang1967,Wang2012,Girardeau2010a,Girardeau2010b,Astrakharchik2008,Girardeau2012,Panfil2013,Lebek2022}, including, for example, the Lieb-Liniger model~\cite{Batchelor2005,Lieb1963b,Franchini2017}, or lattice~\cite{Wang2012} or dipolar systems~\cite{Astrakharchik2008,Girardeau2012}.

The key aspect underlying this phenomenon lies in the similarity between certain eigenstates of system Hamiltonians before and after the quench. The spectrum of strongly attractive system can be divided into two classes of states~\cite{Durr,Lebek2022,Syrwid2022}. The first class contains deeply bound states and the second one consists of sTG states that resemble eigenstates of the TG system. Crucially, the initial TG state in the quench protocol has a vanishing overlap with the first class of states, hence no collapse is observed in the dynamics. Interestingly, the said structure of the spectrum is already captured by a two-body solution indicating that the appearance of the sTG phase is not a genuine many-body effect but rather a feature of the short-range interaction potential. Accordingly, the sTG states can be studied in experiments with only a few particles~\cite{Zurn2012,Murmann2015}.

\begin{figure}[t!] 
\includegraphics[width=\linewidth]{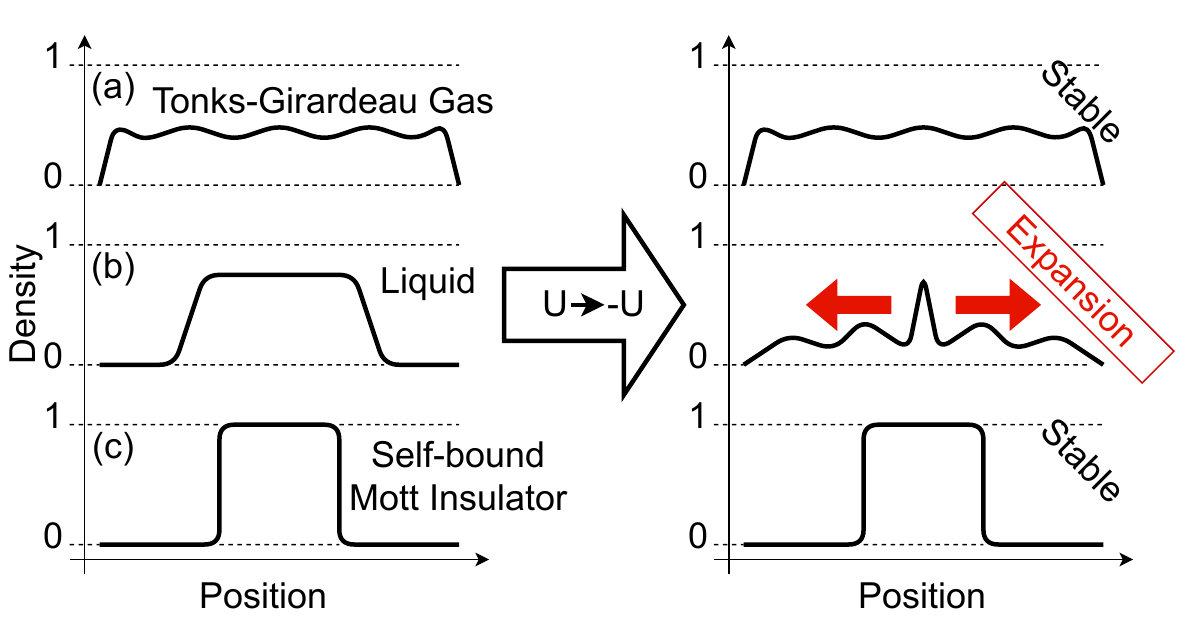}
\centering
 \caption{
 Super-Tonks-Girardeau quench dynamics for the phases of the eBH model: gas and the deeply-self-bound Mott insulator phases are stable, while the liquid droplets expand and eventually evaporate.
 \label{fig:abstract}}
\end{figure}

Finally, a recent experiment with arrays of 1D tubes containing dipolar bosons~\cite{Kao2021}  explored the robustness of the sTG phase against the presence of interactions breaking the integrability of the system.
While the attractive dipolar forces led to diminished stability of sTG gas, dipolar repulsion enhanced the effect. A theoretical explanation of this phenomenon featuring exact diagonalization of three-body systems was proposed in~Ref.~\cite{Chen2023}.

Our work aims to describe sTG quenches in a similar system, in which strong, local interactions are supplemented with weaker non-local interactions. To this end, we consider an extended Bose-Hubbard model (eBH), already accessible in the state-of-art experimental setups both in ultracold gases \cite{Baier201} and solid-state systems \cite{Lagoin2022Sep, Wang2022}. Adding long-range potential to the standard Hubbard model enables studying, e.g., insulating ordered phases at fractional lattice filling \cite{Batrouni1995, Lagoin2022Sep}, ground state phase diagram with quasi-localisation and unexpected topological phases \cite{Goral2002} or non-equilibrium dynamics~\cite{Dutta_2015}. 

Specifically, it has been recently shown that a system confined to a one-dimensional optical lattice with strong on-site repulsion --- close to the Tonks-Girardeau (TG) limit --- and long-range attraction supports liquid formation in the vicinity of a gas-to-insulator phase transition due to the superexchange processes~\cite{Morera2020}. Significantly, the liquid and insulator are characterized by localized density profiles. Here we probe each of these phases by a quench scenario akin to that in sTG gas, see Fig.~\ref{fig:abstract}. We find that strong correlations of sTG phase can manifest drastically in such protocols, leading to the expansion and destruction of bound states.

The observed effect even more strongly contradicts naive intuition that such a quench to strong attraction should lead to a collapse of the system. Moreover, we identify the microscopic mechanism for such behavior in the exactly solvable case of two particles and confirm it using numerical methods suitable for larger systems. 

This paper is organized as follows: in Sec.~\ref{sec:model}, we introduce the model under study.
Then, in Sec.~\ref{sec:TwoAtoms}, we use analytical methods to determine the ground state of a two-boson system  and to study its behavior after a sudden change of short-range interactions from repulsive to attractive.
In Sec.~\ref{sec:fewBody}, we investigate slightly larger systems using many-body numerical methods, focusing on time dynamics.
After that, in Sec.~\ref{sec:ThermodynamicLimitMF}, we analyze systems in the thermodynamic limit. 
Lastly, Sec.~\ref {sec:FiniteSystemsMF} is devoted to the quench dynamics of self-bound states of systems consisting of dozens of bosons.

\section{Model and phase diagram} \label{sec:model}
We consider $N$ bosons confined in a one-dimensional  lattice. The bosons interact via both on-site and nearest-neighbor interactions. Throughout our paper, we assume the nearest-neighbor forces to be weak and attractive. Regarding on-site forces, we are interested in two limiting cases --- either strong local repulsion or strong local attraction. We first analyze the static properties in both cases to understand later the dynamics of the system after a quench in the on-site interaction strength. Note that after such a quench the two interactions --- the on-site and nearest-neighbor ones, are attractive. 

In all cases, we assume that the system is well described by the following Hamiltonian, generally known as the extended Bose-Hubbard Hamiltonian
\begin{align} \begin{gathered} \label{eq:EBH}
    \hH= -J\!\!\!\sum_{j=\!-\!N_h}^{N_h-1}\left( \hb{j}\hbd{j+1} +h.c. \right) + \frac{U}{2}\!\! \sum_{j=\!-\!N_h}^{N_h} \hat{n}_{j}(\hat{n}_j-1) 
    \\
    +V\!\!\!\sum_{j=\!-\!N_h}^{N_h-1}\hat{n}_{j}\hat{n}_{j+1},
\end{gathered} \end{align}
where $\hb{j}$ ($\hbd{j}$) represents a bosonic annihilation (creation) operator at site $j$, $\hat{n}_j=\hbd{j}\hb{j}$,  $N_h=\floor{N_s/2}$, $N_s$ is a number of lattice sites, $J$ is the tunneling coefficient between the neighboring lattice sites ($J>0$), $U$ characterizes the strength of on-site interactions while $V$ stands for the strength of nearest-neighbor interactions. 

To mark the sign of $U$ in the Hamiltonian  \eqref{eq:EBH} --- the strong on-site repulsion ($U \gg J$) or the strong on-site attraction ($U \ll -J$), we introduce $H^+$ and $H^-$, respectively. Namely, the Hamiltonians $H^-$  and $H^+$ have always the same values of the tunneling and nearest-neighbor interaction parameters $J$ and $V$, but the on-site interaction parameters $U$ have opposite signs.

The eBH in the regime of strong, local interactions has a non-trivial ground state phase diagram. Already in the absence of nearest-neighbor interactions $V=0$, the ground state can be one of three different phases: bright soliton ($U<0$), superfluid ($J\gg U>0$), or a lattice equivalent of the Tonks-Girardeau gas ($U\gg J>0$)~\cite{Bloch2008}. 

As shown in Ref.~\cite{Morera2023}, the addition of nearest-neighbor interactions to the eBH with strong local interactions gives rise to new phases called the self-bound liquid and the self-bound Mott insulator (bMI). 
In both phases, the ground state energy per particle is lower than the one of the ideal gas, and the corresponding density profiles are localized in position space, therefore the name ``self-bound".
Moreover, in both phases density profile has a flat top [see Fig.~\ref{fig:abstract}(b,c)] -- similar to quantum droplets thoroughly investigated in the 1D Bose gas~\cite{Oldziejewski20,Petrov2016}.
In the bMI phase, unlike in the liquid one, the lattice sites are either empty or occupied by one atom, practically without any fluctuations (as in the famous Mott insulator phase).

Here, we focus on strong on-site interaction ($|U| \gg J,|V|$) and consider sTG quenches from $H^+$ to $H^-$.
When $V=0$, the situation is clear -- there is only one phase (lattice TG-like gas) and an interaction quench yields standard behavior familiar from continous models~\cite{Astrakharchik2005}. The gas does not evolve significantly and reaches metastable sTG state.
In the case $V \neq 0$, which is the main focus of our paper, we argue that as long as the system starts from a state close to the TG state or bMI, the density profile is stable after the quench. 
The system initially prepared in the liquid phase displays a counter-intuitive behavior after the quench. Despite the attractive character of all forces, the liquid starts to evaporate and does not collapse, as it could be naively expected.
To present and understand these phenomena, we start with the analytical results for two atoms. From that point on, we build the understanding towards systems with larger numbers of particles.

\section{Two atoms} \label{sec:TwoAtoms}

For two bosons, one can find the eigenstates and eigenenergies of the eBH analytically~\cite{Valiente2009,Cifuentes2010}. We describe here the relevant derivation steps and use the results to understand the possible outcomes of the sTG quench. Here, we use the position representation
\begin{align} \label{eq:expanded2atoms}
    \ket{\Psi} = \sum_{j,j'=-\infty}^\infty \Psi(j,j') \ket{j,j'},
\end{align}
where the sum is over the lattice sites. We assume an infinite lattice, therefore the motion of the center of mass $R=(j+j')\,d/2$ and of the distance between two atoms $r=(j-j')d$ are decoupled, with $d$ being the lattice spacing. As a consequence, every eigenstate is a product $\Psi(j,j') = e^{i K R}\psi_K(r)$, where $K$ corresponds to the center-of-mass momentum. 
Substituting this form to the Schr{\"o}dinger equation $\hH \ket{\Psi} = E_K\ket{\Psi}$, one obtains the following equation
\begin{multline}
    -J_K \left[ \psi_K(r-d)+\psi_K(r+d)\right]+\\
    +[U \delta_{r,0}+V(\delta_{r,d}+\delta_{r,-d})
    -E_K]\psi_K(r)=0,
    \label{eq:diff_eq_2atoms}
\end{multline}
where $J_K=2J\cos (Kd/2)$ and $E_K$ is energy of the state.
The difference equation~\eqref{eq:diff_eq_2atoms} can be solved to find the analytical formulas for the scattering and self-bound states~\cite{Cifuentes2010}. 

When considering the system ground state, one can focus on the case where the center-of-mass momentum is zero. The $K=0$ spectrum is schematically presented in Fig.~\ref{fig:two_atoms} for both strong repulsive (a) and attractive (b) on-site interactions. In this figure, the shaded bands correspond to scattering states, while the solid and dashed lines -- to the self-bound states. To explore the parameter space of the model, we fix $U$ and vary the nearest-neighbor attraction $V$. One can see that the number of self-bound solutions and the character of the ground state depends on $V$. In general, there are at most two bound states. Moreover, there is a critical value of the nearest-neighbor attraction,
\begin{equation}
    V_c=-\frac{2JU}{U+4J},
\end{equation} for which the second bound state energy goes below the continuum spectrum of the scattering states. Already here we indicate, that the parameter $V_c$ will be crucial for our further analysis, and in analogy to $H^{\pm}$, we introduce the notation $V_c^{\pm}$. In the following subsection, we analyze the consequences of these facts regarding the outcome of sTG quench. As the ground state for $U \gg 1$ (which is the initial state for the sTG quench) may be either a scattering state or a bound state, we divide the analysis into two parts, finding drastic differences between the two.

\subsection{Super-Tonks-Girardeau quench from a scattering state}
For strong on-site repulsion and sufficiently weak nearest-neighbor attraction (namely $0\leq|V| <|V_c^{+}|$), the ground state is a scattering one [region I in Fig.~\ref{fig:two_atoms}(a)]. Let us now inspect the properties of scattering wave functions to understand the behavior of the system in the $U \to -U$ sTG quench.

The two-body scattering states have the following form
\begin{subequations}
\label{eq:2atoms-solution}
\begin{equation}
    \psi_{K,k}(r \neq 0) = \cos \left(k|r|+\delta_{K,k}\right)
\end{equation}
\begin{equation}
 \psi_{K,k}(0) = \cos  \delta^{(0)}_{K,k}\frac{\cos(kd+\delta_{K,k})}{\cos (kd+\delta^{(0)}_{K,k})}
\end{equation}
\end{subequations}
where the phase shifts read
\begin{subequations}
\begin{equation}
    \tan\delta_{K,k} =  \frac{J_K + V \cos(kd) + \frac{2J_KV}{U}\cos^2(kd) }{ V\sin(kd)-\frac{2J^2_K-2J_KV\cos(kd)}{U}\sin(kd)},
\end{equation}
and
\begin{equation}
    \tan \delta^{(0)}_{K,k}=-\frac{U}{2J_K \sin (kd)}.
\end{equation}
\end{subequations}
The quasi-momenta $k$ are continuous and belong to the interval $[0,\pi/d]$. The energies of the scattering states are equal to $E_{K,k} = -2J_K \cos(kd)$. 

The solution~\eqref{eq:2atoms-solution} is valid for any values of the parameters $U$, $V$, and $J$. In the cases we focus on (i.e., $|U|\gg J$, $|V|$), the phase shift $\delta_{K,k}$ becomes independent from the on-site interactions $\lim_{U \to \pm \infty} \tan\delta_{K,k} =(J_K+V \cos(kd))/(V \sin (kd))$ and $\lim_{U \to \pm \infty}\delta_{K,k}^{(0)}=\mp \pi/2$. 
Consequently, the scattering states in both cases, for extremely strong on-site repulsion and attraction, are equal to each other.
In particular, it means the ground state at $U\to\infty$ is equal to a highly excited eigenstate of the system with $U\to-\infty$. Thus, a quench from $U=\infty \to -U$ has no effect.

From the expansion in $J/|U|$, we see that the situation should remain similar, when one considers finite but large on-site interactions. In this case, the phase shifts are $\delta_{K,k}^{(0)}\approx \mp \pi/2 \pm 2 J_K \sin(kd)/|U|$ and $\tan\delta_{K,k} \approx (J_K+V \cos(kd))/(V \sin (kd)) \pm 2J_K^3/(V^2|U|)$, where $``+"$ corresponds to the solution with a positive $U$.

The similarity between the ground state of $H^+$ ($\ket{H^+_0}$), which corresponds to $k=0$, and one of the excited states of $H^-$ is related to the above-mentioned sTG effect.
The post-quench stability is due to the presence of an excited state of $H^-$, which is highly similar to $\ket{H^+_0}$ (later called the {\it superpartner}).
After quenching, $\ket{H^+_0}$ is an excited eigenstate of $H^-$ and the time evolution leads to changes in the global phase only.

The scattering state property discussed above is consistent with the predictions based on the well-known feature of systems with local interactions only~\cite{Batchelor2005}. However, our system is richer and one can ask a much more interesting question --- will a similar phenomenon also occur for stronger attractive nearest-neighbor interactions, where the ground state of the system is self-bound?

\subsection{Super-Tonks-Girardeau quench from self-bound state}
\begin{figure} 
\includegraphics[width=0.995\linewidth,right]{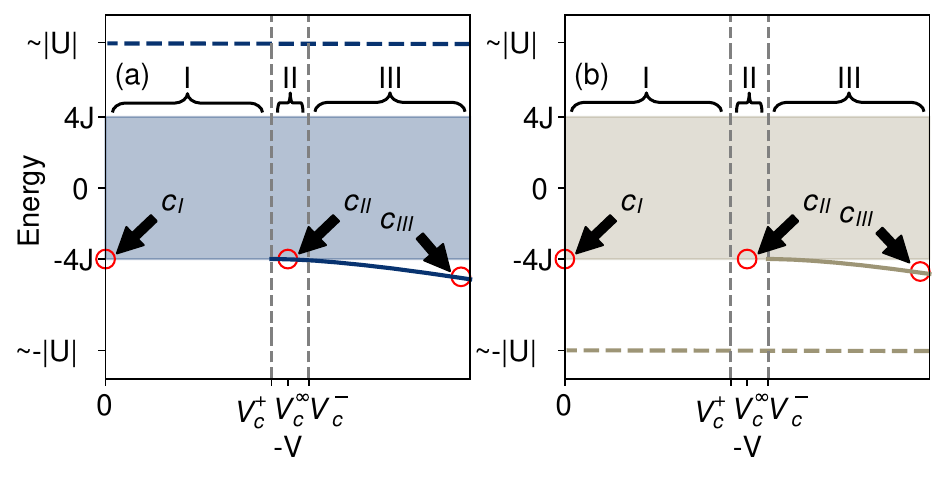}
\centering
\includegraphics[width=0.94\linewidth,right]{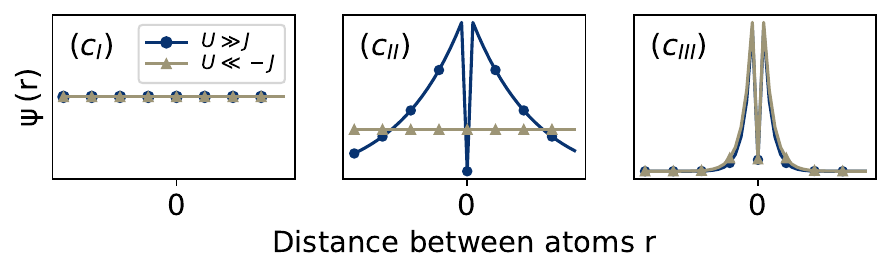}
\includegraphics[width=\linewidth]{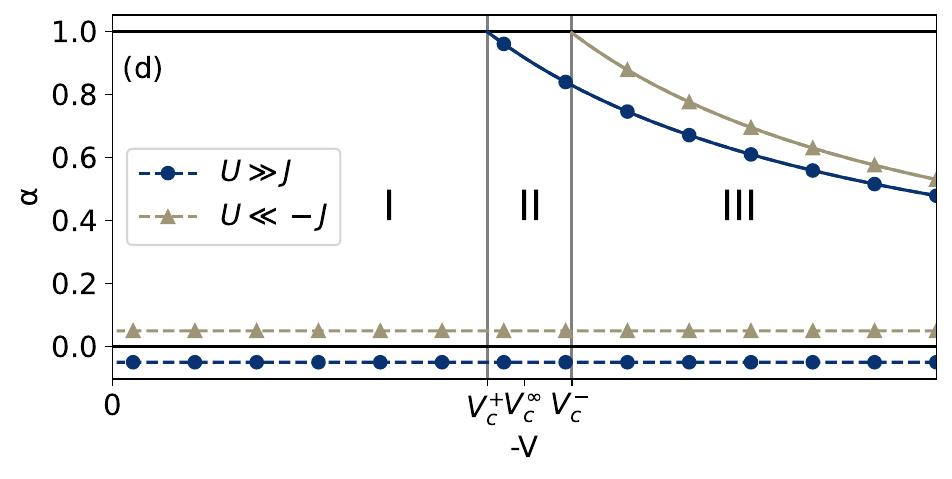}
\centering
\caption{
The upper panels show the $K=0$ energy spectrum of the eBH model  \eqref{eq:EBH} for two atoms in an infinite lattice as a function of the nearest-neighbor interaction strength $V$ for (a) $U\gg J$
and (b) $U\ll-J$. The energies correspond either to the continuum spectrum of the scattering states (the shaded bands) or to self-bound states (solid and dashed lines). There are three relevant regions of the coupling $V$: in (I), the ground state of $H^+$ is a scattering state, in contrast to (II) and (III), where the ground state of $H^+$ is the self-bound state. The difference between (II) and (III) is that in the latter, there exists an excited, self-bound eigenstate of $H^-$ with energy close to the energy of a ground state of $H^+$. \newline
The panels in the middle show three pairs of eigenstates for three different values of nonlocal attraction $V$ corresponding to aforementioned regions (marked in the upper panels by $c_I$, $c_{II}$ and $c_{III}$). We show the ground states of the Hamiltonians $H^+$ (blue lines with circles) and the eigenstates of $H^-$ (yellow lines with triangles) with the energy closest to the energy of $H^+$ GS. \newline
The bottom panel (d) shows the coefficients $\alpha$ characterizing the two-body self-bound eigenstates. 
In region II there is no self-bound eigenstate of $H^-$ (solid yellow lines with triangles) with $\alpha$ similar to that of $H^+$.\label{fig:two_atoms} 
}
\end{figure}

All two-body self-bound eigenstates of the Hamiltonian \eqref{eq:EBH} are in the form of exponentially decaying wave functions~\cite{Valiente2009}
\begin{align}\begin{cases}
    \psi_K(0) = \mathcal{N} \frac{4\alpha_K}{U\alpha_K+2(\alpha_K^2+1)} 
    \\
    \psi_K(r_i\neq 0) = \mathcal{N}\alpha_K^{|i|-1}
\end{cases}\end{align} 
with energies $E_K = -J_K(1+\alpha_K^2)/\alpha_K$. By $\mathcal{N}$ we denote the normalization constant.
The base of the exponent, $\alpha_K$, is determined by the Hamiltonian parameters and can be easily obtained by substituting Eq.~\eqref{eq:expanded2atoms} into the Schr{\"o}dinger equation.
This substitution leads to
\begin{align} \label{eq:2atoms_self_bound}
    J_K V \alpha_K^3 + (UV\!-\!J_K^2)\alpha_K^2 +J_K(U\!+\!V)\alpha_K \!+\!J_K^2 = 0.
\end{align}
The resulting eigenenergies of the self-bound states are depicted with lines in Fig.~\ref{fig:two_atoms}(a),(b), whereas the corresponding coefficients $\alpha_K$ satisfying this equation (with an additional condition $|\alpha_K|<1$, to guarantee the decay of wave function) are shown in  Fig.~\ref{fig:two_atoms}(d). We consider $K=0$ and for brevity, we omit the index $K$, i.e., $\alpha \equiv \alpha_{K=0}$ in there.

Looking at Fig.~\ref{fig:two_atoms}(d), one can see that the values of the coefficients $\alpha$ form two distinct pairs of branches --- the lower pair existing for the whole range of $V$ and the upper one, that appears for $|V| \leq |V_c|$.
The lower pair corresponds to a strongly localized state with energy $E\approx U$. The upper branch is particularly relevant to our considerations, as for repulsive interactions $U>0$ \textit{it corresponds to a ground state of the system} [see Fig.~\ref{fig:two_atoms}(a)], and for attractive interactions $U<0$ to a certain \textit{excited state with a similar wave function}, as shown in Fig.~\ref{fig:two_atoms}($\rm{c_{III}}$).
Crucially, note that the point of appearance of the upper pair depends on $U$ as
\begin{equation}
    V_c^{\pm}\, \overset{|U|\gg1}{\approx}-2J \pm \frac{8J^2}{|U|}
\end{equation} and for Hamiltonians with large but finite $|U|$ is slightly different for $U<0$ and $U>0$ with the following hierarchy
\begin{equation}
    V_c^-<V_c^\infty<V_c^+,
\end{equation}
where $V_c^\infty=-2J$. This difference implies that for $V_c^-<V<V_c^+$ (region II in Fig.~\ref{fig:two_atoms}) the ground state of $H^+$ has no superpartner among the eigenstates of $H^-$. 
A superpartner appears in region III, and it gradually approaches the branch of $H^+$ states.

With this simple analysis, we can identify three ranges of the nearest-neighbor interaction parameter $V$ with different dynamics after a sTG quench.
Since in regions I and III the ground state $\ket{H^+_0}$ has a superpartner, it will be stable after the quench.
In contrast, we do not expect such stability of the system initiated in the ground state from region II. We show in the next chapters for larger systems that in such a scenario the ground state of $H^+$ is a superposition of scattering eigenstates of $H^-$, so the quench results in an expansion of the atoms that were bound together before the quench.

\begin{figure}
    \centering
    \includegraphics[width=0.5\linewidth]{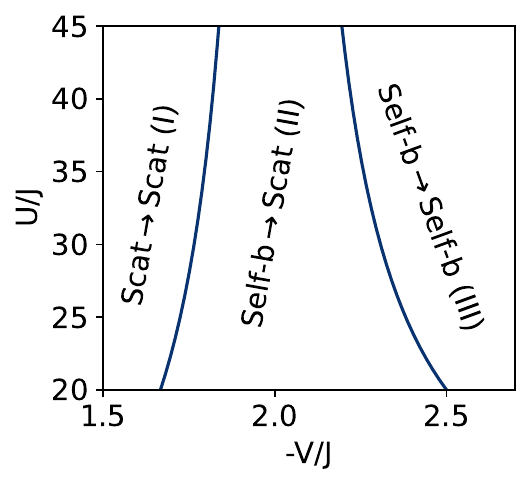}
    \caption{Super-Tonks-Girardeau quench diagram for the two-body system. In region I, the ground state of the system is a scattering state (Scat), which is stable in sTG quench. Region II corresponds to self-bound states (Self-b), which expand after the quench to strong attraction. Finally, in region III the bound ground state survives the change of interactions and is stable after the quench.}
    \label{fig:stg_diagram_2atom}
\end{figure}

The sTG quench diagram is presented in Fig.~\ref{fig:stg_diagram_2atom}.

As we will see later a similar, albeit a little bit richer, diagram appears when one considers many-body systems. In the limit of a large number of particles, where the phase diagram becomes sharply defined~\cite{Morera2023}, the region I will correspond to the gaseous phase, region II will cover liquid and weakly bound Mott insulators, whereas region III will consist solely of deeply-self-bound Mott insulator phase.

The existence of three regions in the sTG quench diagram may be intuitively understood as follows. It is a well known fact, that the sTG state has stronger correlations than the TG system. In particular, it is characterized by a higher pressure than the TG gas (for the same density). When we consider region II, the initial state for the quench, $|H_0^+ \rangle$ is \textit{weakly bound} -- the energy is slightly below the energy of non-interacting system $-4J$ and nearest-neighbor attraction barely compensates local repulsion. When we quench the system to the stronger sTG correlations, we effectively increase the contribution to the ``pressure" from on-site interactions. For weak, nearest-neighbor attraction, i.e., in region II, such increase leads to destruction of the initial, bound state. When we sufficiently increase attraction strength it is harder to destroy the bound state and it survives the quench -- this happens in the region III. We believe that this intuition related to pressure increase from the quench carries over also to larger systems, which we study in the following sections.

Before we turn to truly macroscopic systems, let us make an intermediate step and study few-body systems that are accessible with well-controlled numerical methods.

\section{Few-body systems} \label{sec:fewBody}

\begin{figure} 
\includegraphics[width=\linewidth]{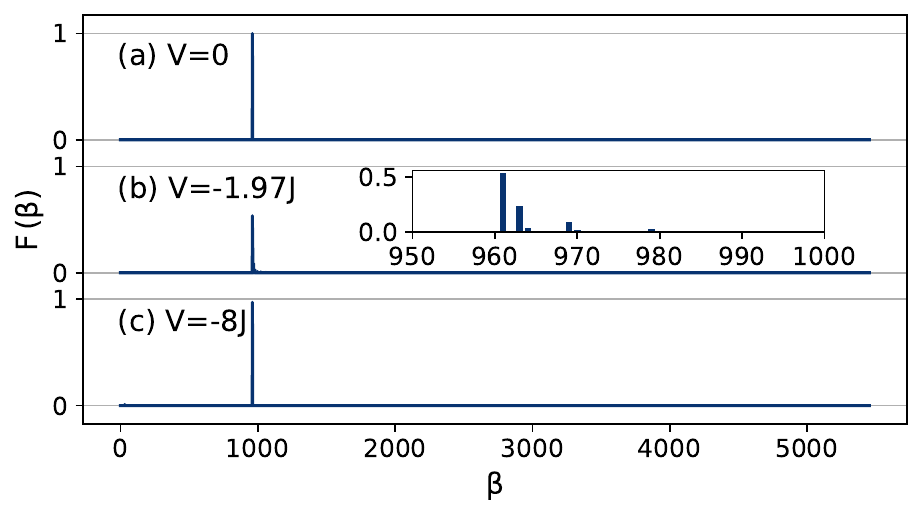}
\centering
\caption{
Fidelity $F(\beta)$ between the ground state of $H^+$ and eigenstates of $H^-$ for $N=3$ atoms in lattice with $N_s=31$ sites, with $|U|=40J$.
For $V=0$ (scattering state) and $V=-8J$ (deeply self-bound state), there is a highly excited eigenstate, the superpartner, practically equal to $|H_0^+\rangle$.
In contrast, the ground state of $H^+$ with an intermediate value of $V=-1.97J$ (weakly self-bound) is a superposition of several states with different energies, which makes it unstable after the quench.
\label{fig:overlapsIII}}
\end{figure}
Systems consisting of more than two atoms are far more complex to analyze. 
The many-body problem cannot be solved analytically, and we must invoke numerical methods. For small systems, it is possible to diagonalize Hamiltonian \eqref{eq:EBH} numerically. To analyze the system in the context of sTG quench, we diagonalize the Hamiltonians $H^+$ and $H^-$ and look for superpartners.
In order to do so, we compute the fidelity between the ground state $\ket{H^+}$ and the eigenstates of the $H^-$ 
\begin{equation}
    F(\beta)= |\langle H^+_0|H^-_\beta\rangle|^2,
\end{equation}
where $|H^-_\beta\rangle$ is the $\beta$-th eigenstate of $H^-$.
Importantly, as in the simpler case of two atoms, we find three regions of parameters.
In our calculations, we fix the on-site interaction strength to  $|U|=40J$, whereas the representative nearest-neighbor attraction strengths are here  $V=0$ (region I), $V=-1.97J$ (region II), and $V=-8J$ (region III). Let us note here, that these three examples are few-body precursors of gaseous, liquid and bMI phases, respectively --- we will see in the next section that these particular parameters lead to the representation of all three phases of the phase diagram also for macroscopic systems [see Fig.~\ref{fig:phases_uniform}(a)].
\begin{figure} 
\includegraphics[width=\linewidth]{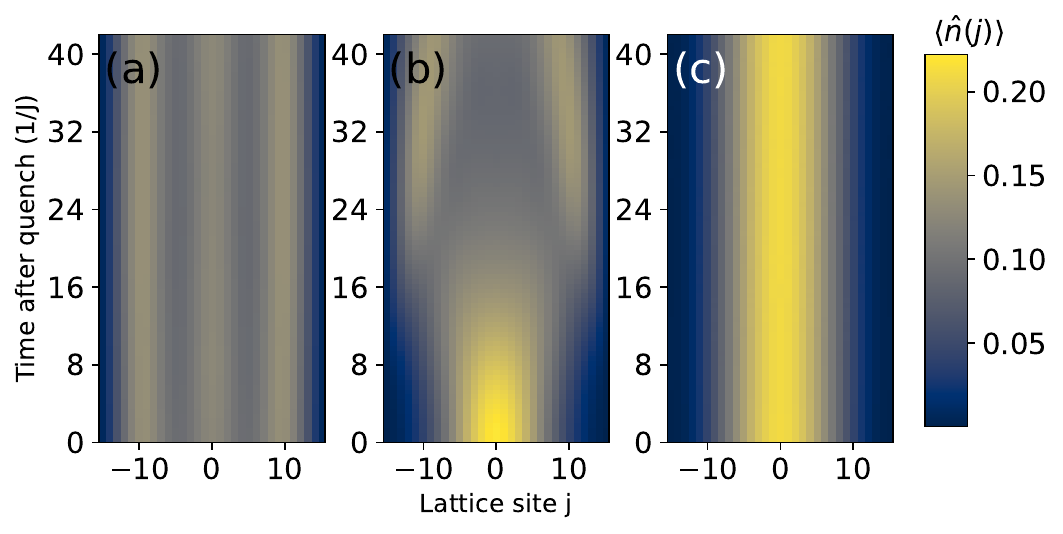}
\includegraphics[width=\linewidth]{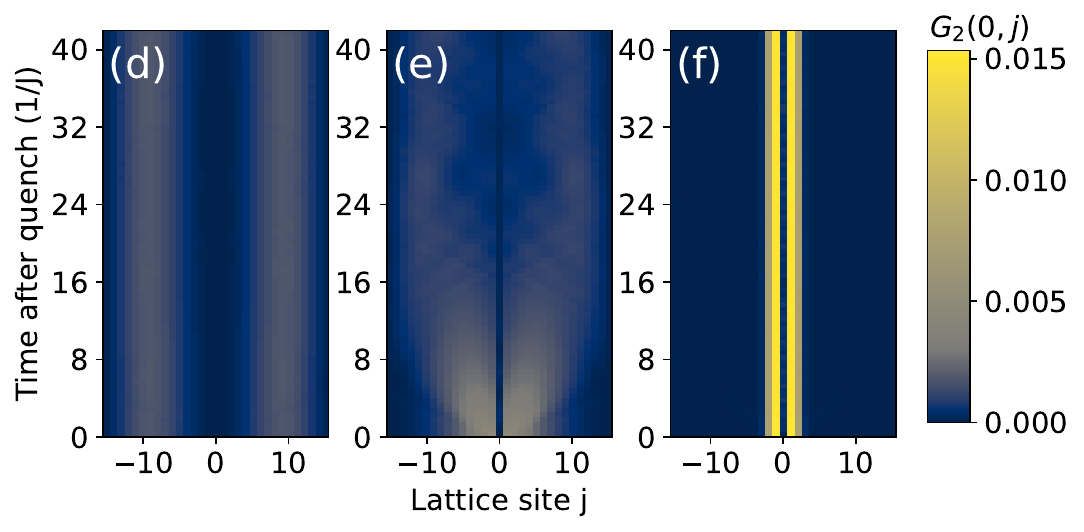}
\centering
\caption{
Density profiles (top panels)  and density-density correlations (bottom panels) as a function of time for the quenched system initiated in the ground state of $H^+$. The nearest-neighbor interaction strengths are (a,d) $V=0$, (b,e) $V=-1.97J$, and (c,f) $V=-8J$.
It can be seen that for the weak (a, d) and very strong (c, f) attraction, both the density profile and the correlations remain unchanged for a long time.
At the same timescale, the weakly bound state  corresponding to $V=-1.97J$ fully evaporates (b, e).
All figures correspond to the number of atoms $N=3$ and $|U|=40J$.
\label{fig:3_tev_nis}}
\end{figure}
We begin our numerical analysis with just three atoms. In Fig.~\ref{fig:overlapsIII} we show fidelities $F(\beta)$ between the ground states for the abovementioned parameters and the eigenstates of the corresponding Hamiltonians $H^-$.
We observe that in the case of the parameters belonging to region I [vide Fig.~\ref{fig:overlapsIII}(a)] and region III [vide Fig.~\ref{fig:overlapsIII}(c)] the maximal fidelity is almost equal to $1$. It means that these states are very similar to some of the excited states of the corresponding $H^-$ Hamiltonians, i.e., they have superpartners. In contrast, the state belonging to region II turns out to be a superposition of many eigenstates of $H^-$ --- exactly as for the two-body system. This suggests that after a quench $U\to-U$, the ground states from regions I and from region III will be stable, whereas the ground state from the region II will not.

\begin{figure} 
\includegraphics[width=\linewidth]{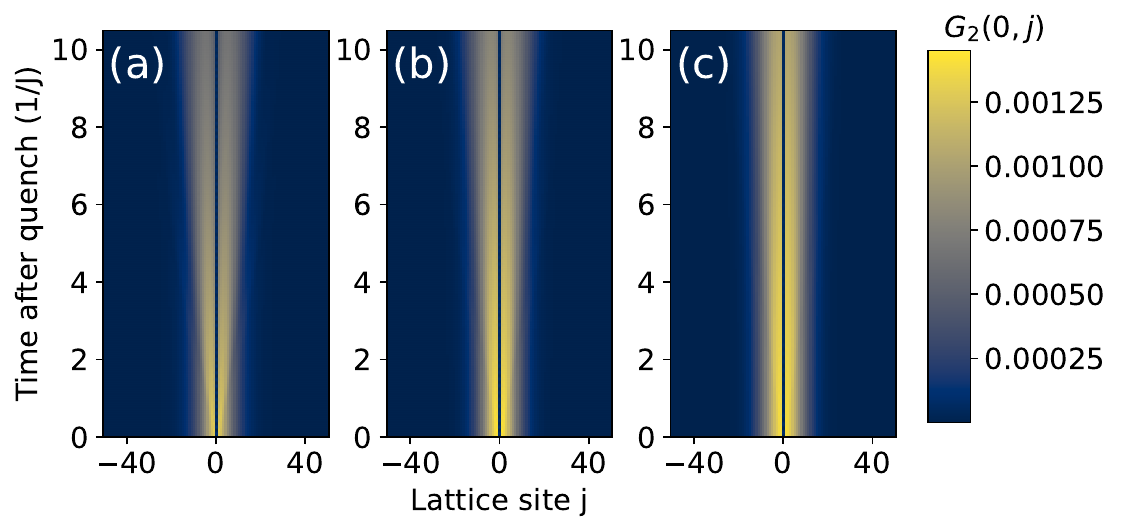}
\centering
\caption{Density-density correlations of a weakly self-bound ground states of $H^+$ in a systems consist of (a) $N=4$, (b) $N=6$ and (c) $N=8$ atoms.
As for the weakly self-bonded ground state of the three-atomic system, also here we can see a rapid increase in the distance between the atoms after quenching.
This observation may be interpreted as a gradual evaporation of the self-bound state.
Here $|U|=40J$ and $V=-1.97J$. The data in this figure were obtained by using \mm{DMRG and TDVP methods \cite{itensor,Yang2020}.}
\label{fig:468_tev}}
\end{figure}
We confirm these predictions based on the spectral analysis by studying the full dynamical problem using the time-dependent variational principle (TDVP) \cite{Yang2020}. We track the dynamics of the density profiles and the second-order correlation function defined as \begin{equation}
    G_2(j,j') = \frac{1}{N^2}\langle \hbd{j}\hbd{j'}\hb{j'}\hb{j}\rangle.
\end{equation}
The time evolution of these quantities after quenching is shown in Fig.~\ref{fig:3_tev_nis}. 
As expected, the systems in regions I and III are insensitive to the quench. This is not the case for the system with parameters from region II [vide Fig.~\ref{fig:3_tev_nis}(b)], in which we observe expansion. Clearly, expansion of the state is visible both on the level of density profile and in the two-body correlation function.
This phenomenon, the expansion of the state despite the drastic increase in the attraction between atoms, can also be observed for a bit larger systems.
For $N=4,6$ and $8$ bosons and parameters from region II, we use DMRG \cite{itensor} to find the corresponding ground states and then the TDVP to study dynamics after the quench.
The results for dynamics of the $G_2$ function, presented in Fig.~\ref{fig:468_tev}, show qualitatively the same features as dynamics for $N=3$ atoms --- the quenched system seems to evaporate.

The number of atoms and the time of evolution are very limited due to numerical difficulties. 
After a quench, the GS of $H^+$ becomes a highly excited eigenstate of $H^-$ (if there is a superpartner) or a superposition of some highly excited eigenstates (if there is not). As a consequence, the entanglement of a system grows rapidly during evolution, making the numerically  accessible time scales not sufficiently long to distinguish stable systems from evaporating ones.
Nevertheless, examples provided here confirm that existence of three regions of parameters defining the outcome of sTG quench is not only a feature of two-body problem, but is true for larger systems.

We treat few-body systems as an intermediate step and do not prepare a sTG quench diagram akin to that presented in Fig.~\ref{fig:stg_diagram_2atom}. Moreover, we expect that precise boundaries between these three regions will change as $N$ is increased. Conjectural form of such a diagram in thermodynamic limit will be presented in the next section.
\section{Perturbative analysis for macroscopic systems\label{sec:ThermodynamicLimitMF}}
In this section we aim to describe macroscopic systems, for which numerical methods used for few-body systems fail due to the large Hilbert space dimension. We do not have access both to the many-body spectrum and to the exact quench dynamics. However, we provide some understanding using approximate, perturbative description of the system developed in a very similar setting~\cite{Morera2023}.

Since we consider a system of highly repellent bosons ($U \gg J$), the system governed by Hamiltonian \eqref{eq:EBH} can be effectively mapped to the fermionic one, with effective Hamiltonian $\hH_{\rm eff}$ \cite{Cazalia03}
\begin{align} \begin{gathered} \label{eq:Heff}
    \hH_{\rm eff} = -J\!\!\!\sum_{j=-N_h}^{N_h-1} \!\left( \hcd{j}\hc{j+1} +h.c. \right)
    -
    \frac{4J^2}{U}\!\!\! \sum_{j=-N_h}^{N_h-1} \hn{j}\hn{j+1}
    \\
    +
    \frac{2J^2}{U}\!\!\!\! \sum_{j=-N_h+1}^{N_h-1} \left( \hcd{j-1}\hn{j}\hc{j+1}\!+\!h.c. \right)
     + 
     V\!\!\!\sum_{j=-N_h}^{N_h-1}\hat{n}_{j}\hat{n}_{j+1},
\end{gathered} \end{align}
where $\hc{j} (\hcd{j})$ represents a fermionic annihilation (creation) operator at site $j$.

\begin{figure} 
\includegraphics[width=\linewidth]{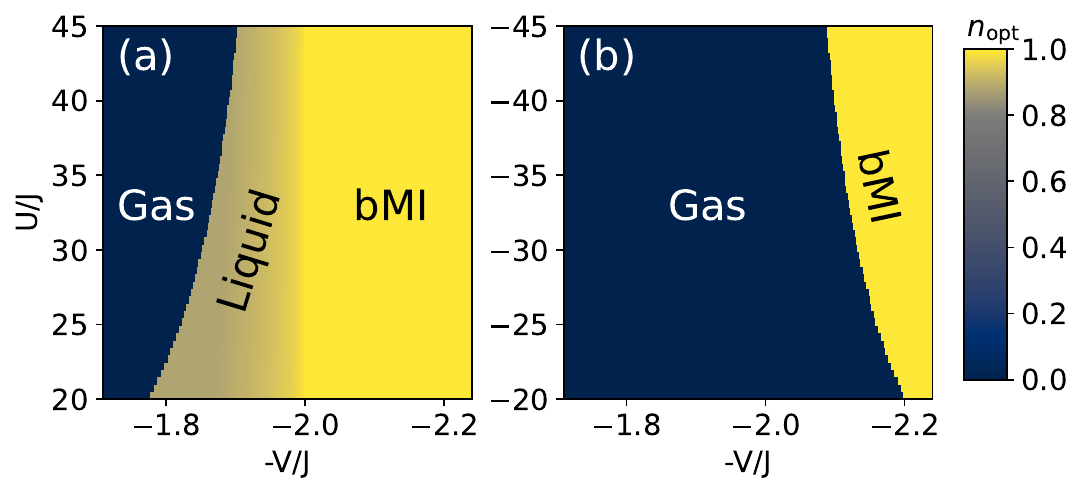}
\includegraphics[width=0.5\linewidth]{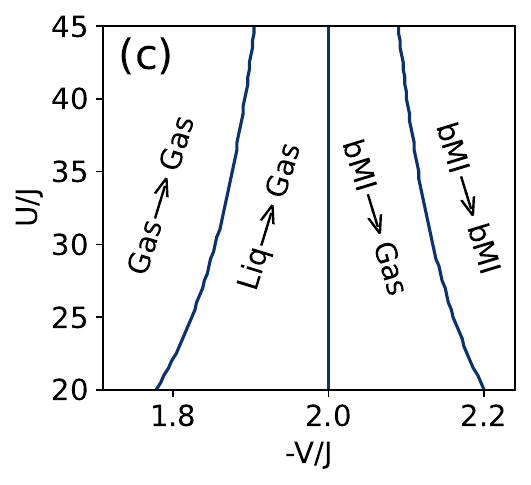}
\centering
\caption{ The figures (a) and (b) present the gas density at which the energy functional \eqref{eq:uniform_energy} reaches a minimum ($n_{\rm opt}$).
There are three possible states: gas ($n_{\rm opt}=0$), liquid ($0<n_{\rm opt}<1$), and a bMI ($n_{\rm opt}=1$).
Note that the liquid appears only for positive values of $U$ (left picture).
If the sign of on-site interactions is reversed (right image) the liquid and a part of the bMI are replaced by gas.
This leads us to the conclusion that a sudden change of parameters from $U\gg J$ to $-U$ causes the evaporation of the liquid to the gaseous state. 
\label{fig:phases_uniform}}
\end{figure}

The energy of such a system can be obtained perturbatively, using the ground state of the non-interacting lattice fermions as the starting point (see \cite{Morera2023} for more details). In thermodynamic limit, one can estimate the ground state energy of the system by considering the expectation value of \eqref{eq:Heff} in the ground state of lattice Fermi gas. The resulting energy functional reads
\begin{align} \begin{gathered} \label{eq:uniform_energy}
    E(n)/N  = - 2J \frac{\sin(\pi n)}{\pi n} - \frac{4J^2n}{U} \left( 1 - \frac{\sin(2\pi n)}{2\pi n} \right) \\+ V n\left( 1 - \frac{\sin^2(\pi n)}{\pi^2 n^2} \right),
\end{gathered} \end{align}
where $n$ is the global density of atoms in a lattice, which cannot be larger than one (reminiscent of the Pauli exclusion principle).
It was previously demonstrated in Ref.~\cite{Morera2023}, that in a nearly identical system for $|U|\gg |V|$ there are three distinct phases -- the gaseous phase, the liquid, and the self-bound Mott insulator. The difference between the system considered here and that of~\cite{Morera2023} is the range of interactions -- while we consider nearest-neighbor coupling, the Ref.\cite{Morera2023} studied dipolar interactions. However, we find that this difference does not alter the phase diagram in a dramatic way, changing the boundaries but not the phases. The phase diagram of the system is presented in Fig.~\ref{fig:phases_uniform}(a).

\begin{figure} 
\includegraphics[width=\linewidth]{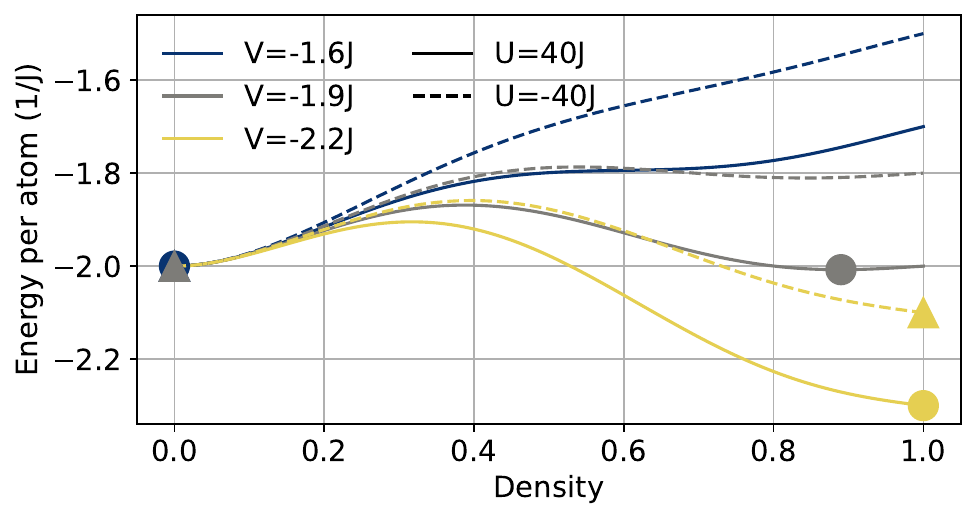}
\centering
\caption{The energy of the dipolar bosons in a one-dimensional lattice for several values of the nearest-neighbor coupling $V$ with a fixed on-site interaction $U=40J$ (solid lines) or $U=-40J$ (dashed lines).
The equilibrium phase of a system is dependent on the optimal density, i.e., the density for which the energy reaches a minimum.
If $n_{\rm opt} =0$ a system finds itself in the gas phase, if $0<n_{\rm opt} <1$ a system is a liquid and for $n_{\rm opt} =1$, it matches the bMI.
It is worth noting that for both $U=40J$ and $-40J$ positions on energy minimum are the same for gas and deeply self-bound Mott insulator, but for liquid, it jumps to $0$.
What may be found is that this effect occurs every time the ground state of $H^+$  is liquid (vide Fig.~\ref{fig:phases_uniform}).
\label{fig:energy_uniform}}
\end{figure}

Following~\cite{Morera2023}, one can identify the phases by studying the energy functional \eqref{eq:uniform_energy}. The functional is minimized with respect to density, which is restricted to the interval $n \in [0,1]$.
If the energy reaches a global minimum for density tending to zero the system is in a gaseous phase -- physically, it is energetically favorable to expand the bosons to fill the whole container.
A totally different situation takes place if the density minimizing the energy $n_{\rm opt}$ approaches the unit filling limit, i.e.,  $n_{\rm opt} = 1$ -- then we expect a bMI phase.
The third phase -- the liquid one -- exists only in a relatively narrow range of nearest-neighbor couplings $V$, where the optimal density is located between zero and one. Interestingly, the appearance of the liquid phase can be attributed to superexchange processes, which yield the $J/U$ correction to the energy~\cite{Morera2023}. Without this term, which effectively introduces \textit{additional attraction}, functional \eqref{eq:uniform_energy} does not predict liquid phase in the system. The boundary between liquid and bMI can be identified from \eqref{eq:uniform_energy} and is defined by the critical value of coupling between adjacent lattice sites:
\begin{equation}
    V_\text{bMI}=-2J.
\end{equation}
All these cases are illustrated with solid lines in Fig.~\ref{fig:energy_uniform} for $U=40J$ and the weak nearest-neighbor attraction equal to  $V=-1.6J$ (gas), $V=-1.9J$ (liquid) and $V=-2.2J$ (bMI).

Eq.~\eqref{eq:uniform_energy} provides a basic understanding of the system with $U>0$ and negative $V$ and perturbatively describes the many-body ground state of $H^+$. 

At this point, the natural question arises: 
how to extend the analysis to the attractive system in our search for superpartners?
Without access to the full spectrum we cannot fully address these questions. Let us note however, that it is natural to conjecture that the eigenstates of strongly attractive system, if the sTG effect occurs, should be described using perturbation theory around fermionized lattice TG gas with $-\frac{J}{|U|}$ taken as a small parameter. In our analysis, we follow this observation. The energy functional that we use for such states is the same as \eqref{eq:uniform_energy}. However, now $U$ is negative and superexchange correction brings \textit{additional repulsion} to the system. As we will see later, this additional repulsion does not allow for liquid formation in the system.

In Fig. \ref{fig:energy_uniform}, in addition to solid lines (describing system for $U=40J$) we plot with dashed lines the energy functional for $U=-40J$, i.e. the energies of the hypothetical superpartners.

One may observe that optimal densities for the weak (gas) and strong (bMI) nearest-neighbor attraction are identical for both, positive and negative $U$.
In contrast, for the intermediate value of nearest-neighbor attraction $V=-1.9J$ the optimal densities differ significantly.
For $U>0$ the optimal density is
$n_{\rm opt}\approx0.9$, (meaning a liquid) whereas for $U<0$, we found $n_{\rm opt}=0$, which indicates the gaseous phase.
This observation suggests that if $|H^+_0\rangle $ has a gaseous or bMI-type character, it may have a superpartner and therefore remains stable after the quench, whereas the liquid should evaporate --- exactly as observed in the previous section devoted to the few-body systems.
To summarize this analysis, we show in Fig.~\ref{fig:phases_uniform}(b) a phase diagram corresponding to hypothetical superpartners.
It is clearly visible there that the entire liquid phase (and a part of the bMI phase) for $U\gg J$ is replaced for $-U$ by a gaseous phase. In other words, we expect that systems with large, negative $U$ do not host liquid-like eigenstates.

In Fig.~\ref{fig:phases_uniform}(c), we mark the speculated scenarios for the dynamics after the quench, depending on the phase one starts with. Note the similarity with the two-body case (cf. Fig.~\ref{fig:stg_diagram_2atom}). However, now we see four regions, instead of just three: two correspond to post-quench stability and two to evaporation due to the quench to strong attraction. This is because of the appearance of an additional phase in the many-body phase diagram -- in the many-body scenario we classify the system as gas, liquid or bMI in contrast to the two-body case, where division into bound and scattering states is sufficient. As we increase $N$ and pass from Fig.~\ref{fig:stg_diagram_2atom} to Fig.~\ref{fig:phases_uniform}(c) the boundaries change and the region in the middle splits into two, both corresponding to evaporation of the localized structures.

In the next section, we extend our analysis to inhomogeneous systems and construct local density approximation (LDA) dynamical theory using energy functional~\eqref{eq:uniform_energy}.

\section{Local density approximation for inhomogenous system\label{sec:FiniteSystemsMF}}
In the previous section, we assumed that the system of strongly repulsive bosons may be described by the effective fermionic Hamiltonian and that the density is uniform in the whole lattice. As a result, we have obtained the energy as a function of a gas density, given by Eq.~\eqref{eq:uniform_energy}. 
This function --- with certain changes --- should also be applicable to large but finite systems.
In order to adapt it, we need to break the translational symmetry, symmetrize the nearest-neighbor interactions, and include the energy of an envelope from the density profile to the final energy.
These assumptions lead us to 
\begin{align} \begin{gathered} \label{eq:meanField1}
E[\bm{c},\bm{c^*}] \!\!=\!\! 
-J\!\!\sum_{k=-N_h}^{N_h-1} \!\!\left( \!c_k c_{k+1}^* \!-\! 2n_k \!+\! c_k^* c_{k+1}  \!+\!2 \frac{\sin(\pi n_k)}{\pi}\!\right) 
    \\- \sum_{k=-N_h}^{N_h}  \frac{4J^2 n_k^2}{U}\left( 1-\frac{\sin(2\pi n_k)}{2\pi n_k}\right)  
    \\
    + V \sum_{k=-N_h}^{N_h-1} n_k n_{k+1} \left( 1 - \frah \frac{\sin^2(\pi n_k)}{\pi^2 n^2_k} -\frah \frac{\sin^2(\pi n_{k+1})}{\pi^2 n^2_{k+1}}\right),
\end{gathered} \end{align}
where $c_k$ is a complex density amplitude in a $k$th node and $n_k = |c_k|^2=\langle \hat{n}_k \rangle$.
Then, the functional \eqref{eq:meanField1} is used to obtain the temporal dependency of coefficients $c_j(t)$~\footnote{We use the energy functional to first construct a  Lagrangian and then solve the Euler-Lagrange equations of motions for variables $c_j(t)$. }
\begin{align} \label{eq:MF_td1}
    i \partj{t} c_j = \partj{c_j^*} E[\bm{c},\bm{c^*}].
\end{align}
which we use to study the dynamics. 
In practice, we also use the functional~\eqref{eq:meanField1} to find the ground state -- we use for that the imaginary time evolution method.

\begin{figure}[t]
\includegraphics[width=\linewidth]{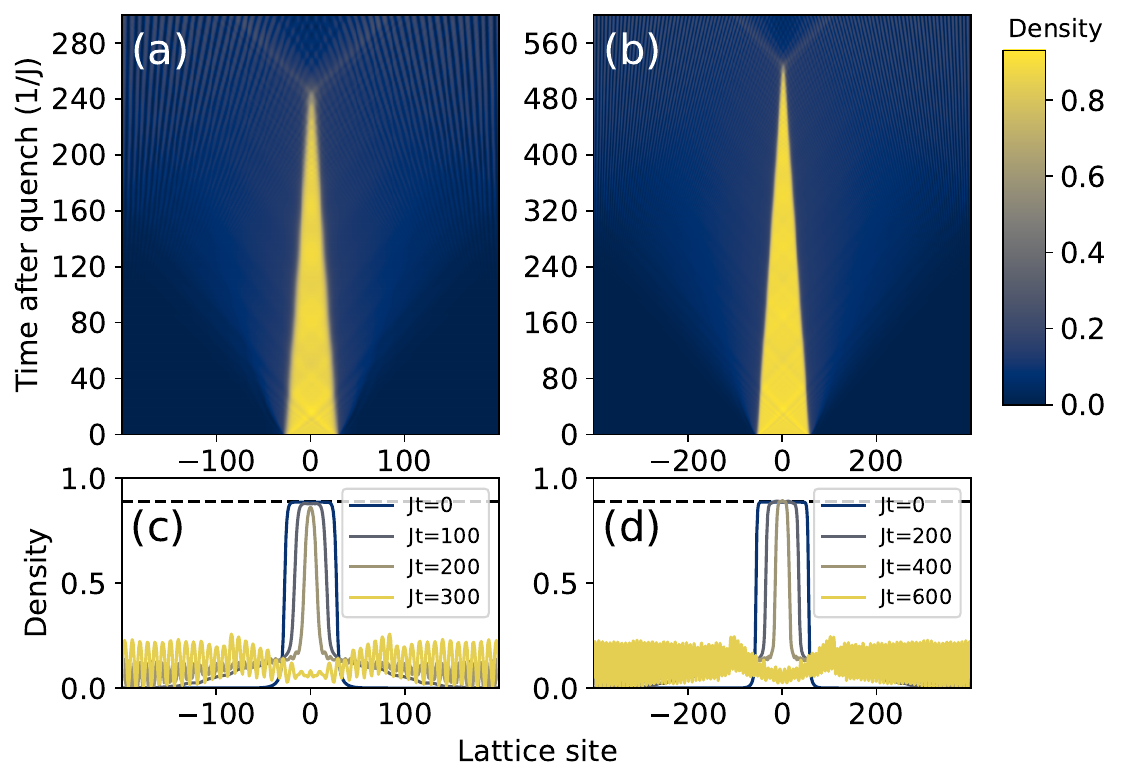}
\centering
\caption{ Dynamics of a droplet with (a, c) $N=50$ (b, d) $N=100$ bosons after a sudden change of on-site interactions from $U=40J$ to $U=-40J$ with $V=-1.9J$.
After quench the droplet gradually evaporates and ends up in the gaseous phase.
The time dependence of densities is described by Eq.~\eqref{eq:MF_td1}. The dashed lines in (c) and (d) represent the optimal density $n_{opt}$ of a uniform system corresponding to aforementioned parameters $J$, $U$, and $V$.
\label{fig:mf_tev}}
\end{figure}

We find that the resulting density profiles are in good agreement with the forecasts based on the analysis from the previous section. 
In particular, whenever we expect a liquid in a uniform system, we observe a droplet-like state with a characteristic flat-top profile, i.e. a constant density in its bulk (vide Fig.~\ref{fig:mf_tev}(c,d) for $Jt=0$), and these densities are well in line with predictions for homogeneous gas (represented by black dotted lines).
Moreover, the evolution of the system after the quench is consistent with the predictions made in the previous sections.
An example is shown in Fig.~\ref{fig:mf_tev}, where we present two droplets with different numbers of atoms. 
After a sudden change in interactions from $U\gg J$ to $-U$, the droplets start to evaporate gradually. 

Obviously, the approach proposed here does not take into account the possibility of forming deeply bound states due to local attraction. We have checked however in Sec.~\ref{sec:fewBody}, that such processes are suppressed in similar quench protocols involving a few-body system. Moreover, the approach has all the limitations related to the mean-field description of strongly interacting systems. Unfortunately, the full quench dynamics for a many-body system is intractable with present numerical methods and we leave this problem for future numerical and experimental studies.

\section{Summary} \label{sec:Conclusions}
In this article, we studied the super-Tonks-Girardeau quench in the extended Bose-Hubbard model. 
In the typical case without the lattice \cite{Astrakharchik2005}, this effect presents a surprising stability of the system prepared in the Tonks-Girardeau state after an interaction quench -- from strong short-range repulsion to strong short-range attraction. An analogous phenomenon is expected for the Bose-Hubbard model for a dilute gas with strong on-site interaction. This stability requires the existence of a superpartner – a highly excited eigenstate of the "attractive" Hamiltonian, which is almost equal to the system ground state before the quench.

In our quest to study systems with competing interaction types and quantum droplets therein, we look at the lattice model extended by the nearest-neighbor attraction.  The ground state of the extended Bose-Hubbard model can be (i) a gas (counter-part of the Tonks-Girardeau state), (ii) a liquid phase, which we expect to be reminiscent of the dipolar quantum droplet, and (iii)  the so-called self-bound Mott insulator~\cite{Morera2023}.
Our main finding is the lack of stability for the system prepared in the liquid phase. Moreover, despite all post-quench interactions being attractive (on-site and nearest-neighbor ones), there is no collapse – the droplet evaporates, and the particles expand over the whole lattice.

To understand this behavior, we first analyzed the eigenstates of the eBH for just two atoms. We have identified three regions of parameters relevant for the sTG quench. In the first region, the ground state of the two-body system is the scattering state, which is stable after the quench. The second region corresponds to weakly bound states that expand after the quench. In the region III, the nearest-neighbor attraction is sufficiently strong so that the bound state survives the quench. We argue that for few and many-body systems the same scenario occurs. Then, the first region corresponds to the gaseous phase, the region II -- to liquid, while the last one to deeply bound Mott insulator. 

We analyzed the two-body case analytically, the intermediate regime of few-particle systems using the exact diagonalization, DMRG  and TDVP, whereas the many-body systems -- using perturbation theory around free Fermi gas.

Our work shows that a sTG quench may lead to richer behavior of the system, when the contact interactions are supplemented with a nearest-neighbor attraction. The expansion of liquid presented here is a novel and interesting consequence of higher-than-TG pressure of sTG gas. Remarkably, it is even more surprising than the original phenomenon, as we observe the expansion of a gas with both short- and long-range attractive interactions.

Another way of looking at our results is through the superexchange correction to the energy. While for positive $U$, i.e. in the pre-quench state these processes stabilize the liquid, when the sign of $U$ is changed the correction drives the evaporation of the phase. This observation points towards studies of sTG quench in other one-dimensional systems, where superexchange corrections play a crucial contribution to the overall energy.

We present the effects within the eBH model -- a well-examined paradigmatic model that can be realized, for instance, in systems of cold dipolar bosons confined in optical lattices. Even in the context of gases, the Bose-Hubbard model represents a simplification, as it ignores long-range tunnelings and, notably, the density-dependent nature of the parameter $U$, as quantified in studies such as \cite{Campbell2006}. This density dependence is linked to higher energy bands, not accounted for in our study.
The dipolar interaction introduces not only nearest-neighbor effects but also density-dependent tunneling \cite{Sowinski2012, Maik_2013}, along with interactions between particles that are farther apart, giving rise to additional phases as discussed in the comprehensive review 
\cite{Dutta_2015}.
The anisotropy of dipolar potential interaction may alter the on-site density profiles, in the extreme case inducing an on-site collapse. This has been  discussed for instance in \cite{Peter_2012} in the context of the possible experimental realization of the phase diagram predicted in~\cite{Lahaye2010triple} in the simplest possible, triple well system.  Moreover, our analysis is restricted to one dimension, typically achieved experimentally through strong transverse trapping of a suitably diluted sample.
As we focus on low density, weak dipolar interactions ($|V|\ll U$), and deep lattices, most of the aforementioned effects are expected to result in minor corrections. Nonetheless, these corrections are worth the attention in planning an experiment realization.

While we have chosen the eBH model due to recent experimental realizations and numerical convenience of lattice models, we believe that the physics presented here should be visible universally in one-dimensional models with both contact and non-local interactions, in continuum and in the lattice. Our work thus provides an important contribution to the understanding of the sTG phase supplemented with long-range interactions. Such systems were recently addressed both in experimental ~\cite{Kao2021} and theoretical~\cite{Chen2023} works. 
Moreover, the results presented here may be useful in the communities working on the eBH or dipolar quantum droplets. The phenomenon can serve as a probe for the phase diagram with gas, liquid and bMI, hopefully in an experiment.

\begin{acknowledgments}
We would like to thank Bruno Juli\'a-D\'iaz, Mariusz Gajda and Marcin Płodzień for inspiring discussions.
M.M., M.Ł., J.K. and K.P. acknowledge support from the (Polish) National
Science Center Grant No. 2019/34/E/ST2/00289.
We acknowledge the assistance from Center for Theoretical Physics of the Polish
Academy of Sciences which is a member of the National Laboratory of Atomic, Molecular and Optical
Physics (KL FAMO).

The data used to create figures can be found at \url{https://gitlab.com/Arawir/super_tonks_girardeau_quench_in_the_extended_bose_hubbard_model}.
\end{acknowledgments}
\bibliography{bibliography.bib}

\end{document}